\documentclass[aps,preprintnumbers,amsmath,amssymb,prd,superscriptaddress,longbibliography,twocolumn]{revtex4-1}
\usepackage{dcolumn}
\usepackage{color}
\usepackage{subfigure}  
\usepackage{feynmp}
\usepackage{feynmp-auto}
\usepackage{pgfplots}

\newcommand{\nc}{\newcommand}
\nc{\beq}{\begin{eqnarray}}
\nc{\eeq}{\end{eqnarray}}
\nc{\scs}{\scriptstyle}
\nc{\setval}{\fmfset{wiggly_len}{3mm} \fmfset{arrow_len}{1.5mm}
	\fmfset{arrow_ang}{13} \fmfset{dash_len}{1.5mm}\fmfpen{0.125mm}
	\fmfset{dot_size}{2thick}}

\usepackage{bm,latexsym,mathrsfs,enumerate,color}
\usepackage[mathcal]{euscript}
\usepackage[breaklinks=true,unicode=true,urlcolor = blue,colorlinks = true,citecolor = blue,linkcolor = blue]{hyperref}
\usepackage{graphicx}
\usepackage{wrapfig}
\usepackage{easyReview}
\usepackage{dsfont}
\usepackage{bbm}
\usepackage[percent]{overpic}
\usepackage{empheq}
\usepackage{tikz}

\usepackage{outlines}

\usetikzlibrary{positioning,shapes.misc,calc,3d,arrows, decorations.pathmorphing,decorations.markings}
\usepackage{tikz-3dplot}

\renewcommand{\vec}[1]{\bm{#1}}

\def\slashchar#1{\setbox0=\hbox{$#1$}           
	\dimen0=\wd0                                 
	\setbox1=\hbox{/} \dimen1=\wd1               
	\ifdim\dimen0>\dimen1                        
	\rlap{\hbox to \dimen0{\hfil/\hfil}}      
	#1                                        
	\else                                        
	\rlap{\hbox to \dimen1{\hfil$#1$\hfil}}   
	/                                         
	\fi}                                         %

\DeclareMathAlphabet\mathbfcal{OMS}{cmsy}{b}{n}

\usepackage{easyReview}
\usepackage[capitalize]{cleveref}

\begin{document}

\title{Bosonization in 2+1 dimensions via Chern-Simons bosonic particle-vortex duality}

\author{O\u{g}uz T\"urker}
\affiliation{Institute for Theoretical Physics and W\"urzburg-Dresden Cluster of Excellence ct.qmat, TU Dresden, 01069 Dresden, Germany}

\author{Jeroen van den Brink}
\affiliation{Institute for Theoretical Physics and W\"urzburg-Dresden Cluster of Excellence ct.qmat, TU Dresden, 01069 Dresden, Germany}
\affiliation{Institute for Theoretical Solid State Physics, IFW Dresden, Helmholtzstr. 20, 01069 Dresden, Germany}

\author{Tobias Meng}
\affiliation{Institute for Theoretical Physics and W\"urzburg-Dresden Cluster of Excellence ct.qmat, TU Dresden, 01069 Dresden, Germany}

\author{Flavio S. Nogueira}
\affiliation{Institute for Theoretical Solid State Physics, IFW Dresden, Helmholtzstr. 20, 01069 Dresden, Germany}

\date{Received \today}

\begin{abstract}
Dualities provide deep insight into physics by relating two seemingly distinct theories.  
Here we consider a duality between lattice fermions and bosons in (2+1) spacetime dimensions, relating free massive Dirac fermions to Abelian Chern-Simons Higgs (ACSH) bosons.
To establish the duality we represent the exact partition function of the lattice fermions in terms of the \emph{writhe} of fermionic worldlines.
On the bosonic side the partition function is expressed in the writhe of the vortex loops of the particle-vortex dual of the ACSH Lagrangian.
In the continuum and scaling limit we show these to be identical.
This result can be understood from the closed fermionic worldlines being direct mappings of the ACSH vortex loops, with the writhe keeping track of particle statistics.
%
\end{abstract} 

\maketitle 

\section{Introduction}


A duality transformation relates two theories that appear to be very different. Such a mapping is particularly useful if a seemingly hard question in one theory duality-transforms into a simple one in another theory.
Duality transformations for example often invert the coupling constant in the dual theory, {thereby transforming strongly interacting models into weakly interacting ones  and vice versa \cite{Savit_RevModPhys.52.453,kleinert1989gauge}. 
In other cases, the transformation of the coupling constant is not a simple inversion, but rather a more complex function of the original one. 
Analyzing the properties of this function typically still allows to obtain results that would be difficult to achieve otherwise. 
A well known example is provided by the two-dimensional Ising model, whose dual model is again an Ising model. 
In this case the self-duality allows an exact determination of the critical temperature by just looking for the fixed point of the 
duality transformation, a result obtained before Onsager derived the exact solution of the model \cite{Kramers-Wannier_PhysRev.60.252}.   

Of particular interest are dualities that implement a transmutation of particle statistics in addition to a mapping of coupling constants. In one-dimensional quantum systems (1+1 dimensions) such transmuting mappings between fermionic theories and bosonic ones are well-known and go under the general name of bosonization. Due to the fermionic sign structure of the wavefunction the situation is however much more complex in higher dimensions. 
In recent years there has been an intense activity surrounding {\it bosonization dualities} in 2+1 dimensions \cite{SEIBERG2016395,Karch_PhysRevX.6.031043,Mross_PhysRevX.7.041016,aharony2017chern,benini2018three,Raghu_PhysRevLett.120.016602,FERREIROS20181,NASTASE2018145,PhysRevX.7.031051,Dutta_2008,PhysRevD.101.076010}, but indeed it has turned out to be very difficult to obtain exact statements. For instance, while the free massive Dirac fermion in 1+1 dimensions can be exactly mapped into a sine-Gordon model with a particular value of the coupling constant 
\cite{zinn2002quantum,witten1984,Stone_doi:10.1142/2436}, a similar statement in 2+1 dimensions is argued to hold only at the infrared 
stable fixed point of the dual bosonic theory. The latter is given in this case by the following Abelian Chern-Simons Higgs (ACSH) Lagrangian, 
\begin{eqnarray}
\label{Eq:CSAHM}
\mathcal{L}&=&\frac{1}{4\pi}\epsilon_{\mu\nu\lambda}a^\mu\partial^\nu a^\lambda+|(\partial_\mu-ia_\mu)\phi|^2
\nonumber\\
&-&m^2|\phi|^2-\frac{\lambda}{2}|\phi|^4.
\end{eqnarray} 
A particularly simple classical soliton solution arises in the form of a 
so-called "self-dual" CS vortices \cite{Jackiw-Weinberg_PhysRevLett.64.2234}, where "self-dual" here
means
the saturation of the Bogomolny bound \cite{manton2004topological} for the energy, achieved for a certain values of the coupling constants, which generally leads to first-order differential equations for the fields rather than second-order ones. The static vortex solution obtained in this way features a nonzero angular momentum, which is a direct consequence of the Chern-Simons (CS) term. It is interesting to note that a nonzero angular momentum for vortices is forbidden in the case of an ordinary Abelian Higgs model featuring a Maxwell term but sans CS term \cite{Julia-Zee_PhysRevD.11.2227}.  But this no-go theorem does not hold for the theory (\ref{Eq:CSAHM}) as due to the CS term the magnetic flux becomes the source of electric fields. A closely related result has been discussed recently in the context of an axion electrodynamics of vortices for a superconductor-topological insulator structure \cite{Nogueira-Nussinov-van-den-Brink_PhysRevLett.121.227001}.  The remarkable fact is that  for the case of a CS coupling as given in Eq. (\ref{Eq:CSAHM}), the total angular momentum associated to the electromagnetic field and vortex is quantized in units of {$\hbar/2$}
which implies that the CS term transmutes the vortex into a fermion. 
Thus this soliton solution of the classical field equations already suggests a boson-fermion transmutation required by bosonization techniques. 

Historically boson-fermion transmutation within a bosonization framework in 2+1 dimensions has been first discussed by Polyakov \cite{Polyakov_doi:10.1142/S0217732388000398} for a model closely related to (\ref{Eq:CSAHM}), namely, the CP$^{N-1}$ model with a CS term. Polyakov's approach has been elaborated further in Refs. \cite{Hashimoto1989,SHAJI1990,Grundberg1990} and provides an early instance where the mapping of bosons to free massive Dirac fermions in 2+1 dimensions is discussed.   

One of the main results of this paper is to establish a correspondence between free massive Dirac lattice fermions and the bosonic particle-vortex duality of Lagrangian 
(\ref{Eq:CSAHM}) in the continuum limit  using an exact representation of  the partition function of Wilson fermions in terms of the \emph{writhe} associated with the fermionic particle worldlines.
Particle-vortex dualities for the Abelian Higgs model  (sans CS term) in 2+1 dimensions are well established in several different, but closely related approaches  \cite{Peskin1978,THOMAS1978513,Dasgupta-Halperin_PhysRevLett.47.1556,kleinert1982disorder,kleinert1989gauge}: it consists in mapping the worldline of a {\it particle} in a system with global $U(1)$ symmetry (for instance, the XY model) to {\it vortex} lines of the Abelian Higgs model. This particle-vortex duality does not involve a change of statistics, as both sides of the duality involve bosonic fields only. But it suggests a pathway to establish the bosonization duality in 2+1 dimensions: mapping the worldline of free massive fermions to the vortex loops of the Abelian Chern-Simons Higgs (ACSH) model (\ref{Eq:CSAHM}). Interestingly, the lattice form of the ACSH Lagrangian (\ref{Eq:CSAHM}) can be mapped by means of an {\it exact} duality to a Lagrangian of almost the same form \cite{REY1991897}, differing from the original Lagrangian by the presence of a Maxwell term.  

The bosonization duality in 2+1 dimensions can be established exactly in the ultraviolet (UV) regime \cite{Raghu_PhysRevLett.120.016602} and is assumed to hold only approximately in the infrared regime (IR). This makes it important to establish a correspondence between the bosonic particle-vortex duality and the bosonization duality in a way that is as exact as possible. This is not an obvious task, since  integrating out the bosonic matter fields in Eq. (\ref{Eq:CSAHM}) leads to a self-linking of vortex loops, which is a less obvious occurrence in free massive Dirac fermions, as first realized by Polyakov \cite{Polyakov_doi:10.1142/S0217732388000398}. 

The plan of the paper is as follows. In \cref{Sec:dualbos} we discuss the properties of the effective action of the bosonic 
theory, which prepares us for the identification of closed particle worldlines to vortex loops in later sections.  
In \cref{sec:wilsonfermions} the fermionic sector of the duality will be considered in the lattice using the 
Wilson fermion technique.   Since the theory is Gaussian, the partition function can be obtained exactly in the thermodynamic limit. 
However, in 2+1 dimensions free massive Dirac fermions exhibit a nontrivial topology which can only be unveiled via a subtle path integral 
representation of the fermion determinant, $\det(\gamma^\mu\partial_{\mu}+m_F)$ 
\cite{Polyakov_doi:10.1142/S0217732388000398,Goldman2018}.  In fact, despite the absence of interactions with a gauge field, the topologically 
nontrivial feature associated to the parity anomaly is already apparent from a straightforward exact calculation of the current correlation function.  
On the lattice we determine the fermion partition function  by means of a hopping parameter expansion in a way similar to Ref. \cite{Stamatescu1982}.  
This allows an exact representation of the  partition function as a summation over the weights of the all possible fermion worldlines, which are closed loops characterized by  
the writhe number.  The way the writhe arises here is a consequence of the interplay between 
parity symmetry breaking (due to the mass) and the nontrivial topology of spinors in 2+1 dimensions.  In \cref{sec:convfermiondet} we discuss 
the convergence of the hopping expansion and cast the partition function in a form more appropriate to relate to the bosonic 
dual partition function. The latter is discussed in \cref{sec:pv-duality}, where 
the duality transformation  will be performed exactly on the lattice, where in the dual model the 
CS coupling is inverted. 
The bosonization duality implies that any side of the particle-vortex duality can in 
principle be mapped to the Lagrangian of a free massive Dirac fermion. This fact necessarily constraints the CS coupling to have the 
form given in Eq. (\ref{Eq:CSAHM}).  
We emphasize here the important role of the writhe number that naturally emerges when analyzing vortex loops in CS theories \cite{Forte_RevModPhys.64.193}.  
In \cref{sec:pv-duality} we show that the partition function of the dual ACSH theory is given as summation of the weight of the all possible vortex loop configurations,  where we characterize the weight of the vortex loops in terms of the \emph{writhe} of the loops. Finally, in \cref{sec:comparison} the bosonization duality 
is established in the low energy vortex sector.

 \section{Bosonic particle-vortex duality, and properties of the bosonic continuum actions}
\label{Sec:dualbos}
We begin by analyzing the 
(purely bosonic) particle-vortex duality of the ACSH Lagrangian. 
For a general CS coupling $\theta/\pi$, the duality takes the form, 

\begin{subequations}
\label{Eq:pvduality}
\begin{eqnarray}
\mathcal{L}_{b}&=&\frac{i\theta}{4\pi^2}\epsilon_{\mu\nu\lambda}a^\mu\partial^\nu a^\lambda
+|(\partial_\mu-ia_\mu)\phi|^2
\nonumber\\
&+&m_{\text{B}}^2|\phi|^2+\frac{\lambda}{2}|\phi|^4\label{eq:abh}\\ 
&&~~~~~~~~~~\Updownarrow{\notag}\nonumber
\end{eqnarray}
\begin{eqnarray}
\tilde{\mathcal{L}}_{b}&=&\frac{1}{2e^2}(\epsilon_{\mu\nu\lambda}\partial^\nu b^\lambda)^2
+\frac{i\theta_{\text{D}}}{4\pi^2}\epsilon_{\mu\nu\lambda}b^\mu\partial^\nu b^\lambda
\nonumber\\
&+&|(\partial_\mu-ib_\mu)\tilde \phi|^2+\tilde{m}_{\text{B}}^2|\tilde{\phi}|^2+\frac{\tilde{\lambda}}{2}|\tilde{\phi}|^4
\label{eq:vortexabh}
\end{eqnarray}
\end{subequations}
where $\theta_D=-4\pi^4/\theta$. Unlike the free fermionic action 
\begin{align}
\mathcal{L}_{f}&=\bar{\psi}(\gamma^{\mu}\partial_{\mu}+m_{\text{F}})\psi,
\label{eq:fbduality-1}
\end{align}
the bosonic fields in Eqs.~\eqref{Eq:pvduality} are interacting. A duality between Eqs.~\eqref{Eq:pvduality} and \eqref{eq:fbduality-1} can therefore only  hold in a regime in which amplitude fluctuations of the bosonic fields are suppressed. This section establishes when this is the case, and what the properties of the bosonic continuum actions in Eqs.~\eqref{Eq:pvduality} are in this regime. This will be particularly helpful in \cref{sec:pv-duality}, where we rigorously demonstrate the duality in Eqs. (\ref{Eq:pvduality}) on the lattice. The results of this section will allow to clearly identify the parameters of the lattice actions in terms of the parameters of the continuum actions. 

Eqs.~\eqref{Eq:pvduality} subscribes into the context of a standard bosonic particle-vortex duality 
\cite{Peskin1978,THOMAS1978513,Dasgupta-Halperin_PhysRevLett.47.1556,kleinert1982disorder,kleinert1989gauge}, 
which has also been discussed in the presence of topological terms in the past \cite{Cardy-Rabinovici,Cardy-theta,REY1991897}. The dual Lagrangian $\tilde{\mathcal{L}}_b$ features a complex disorder field $\tilde{\phi}$ whose coupling to the gauge field $b_\mu$ leads to 
superconducting vortex lines representing the worldlines of the particles of the Lagrangian $\mathcal{L}_b$.  In the limit $\theta\to\infty$, the bosonic particle-vortex duality  (\ref{Eq:pvduality}) reduces to the well known duality between the $XY$ model 
and a superconductor described by an 
Abelian Higgs model \cite{Peskin1978,Dasgupta-Halperin_PhysRevLett.47.1556,kleinert1982disorder}. On the other hand, for $e^2\to\infty$, 
both Lagrangians have the same form, with the CS term having inverted signs, reflecting the self-duality of the CS 
Abelian Higgs model with its time reversed partner. The gauge coupling $e^2$ in Eq. (\ref{eq:vortexabh}) 
is given by the bare phase stiffness of the Lagrangian of Eq. (\ref{eq:abh}). This statement will be made more precise in \cref{sec:pv-duality}. 

If one adds a Maxwell term 
$(\epsilon_{\mu\nu\lambda}\partial_\nu a_\lambda)^2/(2g^2)$ 
to the Lagrangian (\ref{eq:abh}) 
as an UV regulator, it is well known that for $\theta=0$ the charge-neutral Wilson-Fisher fixed point becomes unstable and that 
the charged IR stable fixed point (sometimes called 
gauged Wilson-Fisher fixed point in the more recent literature \cite{SEIBERG2016395,Goldman2018}) is perturbatively inaccessible for a single complex scalar 
\cite{HLM_PhysRevLett.32.292,LAWRIE19821,Herbut_PhysRevLett.76.4588}. The CS term makes the charged fixed point perturbatively accessible 
provided the RG calculations are performed using a massive scalar field, while there are indications that 
conformality is lost if one studies the RG flow for the critical theory \cite{Nogueira2019}. 

To gain intuition about the role played by the CS term, one can for example compute the properties of the dual model (\ref{eq:vortexabh}) (featuring a regulating Maxwell term) for a fixed uniform scalar field background, $\tilde{\phi}=\tilde{\phi}_0$. Integrating out the gauge field then yields
\begin{eqnarray}
\label{Eq:Ueff-dual}
&&\widehat{U}_{\rm eff}(\tilde \phi_0)=\widetilde m_B^2|\tilde \phi_0|^2+\frac{\widetilde \lambda}{2}|\tilde \phi_0|^4
\nonumber\\
&+&\frac{1}{V}\left\{
\sum_{\alpha=\pm} \ln\det[-\Delta+M_\alpha^2(\tilde{\phi}_0)]
\right.\nonumber\\
&-&\left.\ln\det(-\Delta+2e^2|\tilde{\phi}_0|^2)-\ln\det(-\Delta)\right\}
\nonumber\\
&=&\widetilde m_B^2|\tilde \phi_0|^2+\frac{\widetilde \lambda}{2}|\tilde \phi_0|^4
\nonumber\\
&+&\frac{1}{2\pi^2}\int_{0}^\Lambda dpp^2\left\{\sum_{\alpha=\pm}\ln\left[1+\frac{M_\alpha^2(\tilde{\phi}_0)}{p^2}\right]
\right.\nonumber\\
&-&\left.\ln\left(1+\frac{2e^2|\tilde{\phi}_0|^2}{p^2}\right)
\right\},
\end{eqnarray}
where $V$ is the (infinite) volume, $\Lambda$ is a UV cutoff and, 
\begin{eqnarray}
\label{Eq:Ms}
M_\pm^2(\tilde{\phi}_0)=2e^2|\tilde{\phi}_0|^2+\frac{\delta^2}{2}\pm\frac{|\delta|}{2}
\sqrt{\delta^2+8e^2|\tilde{\phi}_0|^2},
\end{eqnarray}
where $\delta=e^2\theta_D/(2\pi^2)$. 
A Landau expansion of the above effective potential up to $|\tilde{\phi}_0|^4$ yields, 
 \begin{eqnarray}
 \widetilde{U}_{\rm eff}(\tilde \phi_0)&\approx& \left(\tilde{m}_B^2+\frac{\Lambda e^2}{\pi^2}
 -\frac{3\sqrt{2}e^4}{|\theta|}\right)|\tilde \phi_0|^2
 \nonumber\\
 &+&\frac{\sqrt{2}e}{3\pi}|\tilde{\phi}_0|^3+\frac{\tilde{\lambda}}{2}|\tilde \phi_0|^4,
 \end{eqnarray}
 implying that the one-loop photon bubble diagram at zero external momenta gives no correction to the renormalized coupling 
 $\widetilde{\lambda}_R=\widetilde{\lambda}+({\rm quantum~ corrections})$ \cite{Nogueira2019}. 
 This is in stark contrast with the theory where a CS 
 term is absent, where the same diagram is IR divergent and thus cannot be evaluated for zero external momenta 
 \cite{Herbut_PhysRevLett.76.4588}. This fact creates a difficulty to smoothly interpolate between the Abelian CS Higgs model and the 
 standard Abelian Higgs model \cite{Nogueira2019}.    
 Given these considerations, in order to avoid the difficulties associated to the 
critical theory, we will assume in this paper that the fixed point structure is governed by a theory with a nonzero renormalized mass $m_R$ 
for the theory (\ref{eq:abh}) [or $\widetilde{m}_R$ for the theory (\ref{eq:vortexabh})]   
and that the IR fixed point is approached as $m_R\to 0$. 

An interesting question is the role of amplitude fluctuations in Eq.~\eqref{eq:abh}. If the scalar field  $\phi_0$ were in a fixed, homogenous configuration, the analogue effective potential $U_{\rm eff}(\phi_0)$ could easily be obtained from  
(\ref{Eq:Ueff-dual}) by replacing  the background field $\tilde{\phi}_0$ there by $\phi_0$,  
$\theta_D$ by $\theta$ and letting $e^2\to\infty$. We then obtain, 
\begin{eqnarray}
\label{Eq:Ueff}
&&U_{\rm eff}(\phi_0)=m_B^2|\phi_0|^2+\frac{\lambda}{2}|\phi_0|^4
\nonumber\\
&+&\frac{1}{2\pi^2}\int_{0}^\Lambda dpp^2\ln\left[1+\frac{M^2(\phi_0)}{p^2}\right],
\end{eqnarray}
where $M^2(\phi_0)=16\pi^4|\phi_0|^4/\theta^2$. 
By performing the integral explicitly and assuming $\Lambda\gg|M|$, we obtain, 
\begin{eqnarray}
\label{Eq:Ueff-1}
U_{\rm eff}(\phi_0)&\approx& m_B^2|\phi_0|^2+\frac{1}{2}\left(\lambda+\frac{16\pi^2\Lambda}{\theta^2}\right)|\phi_0|^4
\nonumber\\
&-&\frac{32\pi^5}{3|\theta|^3}|\phi_0|^6+\mathcal{O}\left(\frac{1}{\Lambda}\right).
\end{eqnarray}
Thus, in contrast with the standard Higgs model in 2+1 dimensions (i.e., with a Maxwell term and without a CS term) 
\cite{HLM_PhysRevLett.32.292}, the effective potential above  
appears to be unstable (unbounded below) due to the generation of a negative $|\phi_0|^6$. However, since its coefficient is dimensionless and 
thus independent of the cutoff, it can be safely neglected. A more elaborate argument would be to notice that quite generally for a local  
field theory with no more than two derivatives in the classical action a $|\phi|^6$ term is an irrelevant operator, with the corresponding 
coupling constant flowing to zero anyway. Hence, we could easily absorb the constant term into a bare $|\phi|^6$ and set it to zero. 

Thus, after dealing with the negative $|\phi|^6$ contribution, we see that the effective potential resembles a standard Landau theory, with  
$\lambda$ receiving a large shift proportional to the UV cutoff $\Lambda$. We see that even if one starts with $\lambda=0$, a scalar 
field self-interaction $\sim\Lambda/\theta^2$ is generated. We find therefore that the value of the order parameter corresponding 
to the minimum of the the effective potential  (\ref{Eq:Ueff-1}) is attained for $m_B^2<0$ and depends on $\theta$, 
\begin{equation}
\label{Eq:phi-mean-field}
|\phi_{0,{\rm min}}|^2=-\frac{m_B^2\theta^2}{16\pi^2\Lambda+\theta^2\lambda},
\end{equation}
which trivially reduces to the usual mean-field Landau theory result for $|\theta|\to\infty$. 
Note that one can use the UV scale $\Lambda$ to define dimensionless quantities out of both bare parameters $m_B^2$ and $\lambda$, as is 
customary in RG theory \cite{zinn2002quantum}. Thus, we would have $\lambda=\Lambda \hat{\lambda}$, where $\hat{\lambda}$ is dimensionless. 
The coupling constant $\lambda$ can only be disregarded in (\ref{Eq:phi-mean-field}) if $\theta^2\hat{\lambda}\ll 16\pi^2$.  

So far we have not considered the vortices of the theory, which are connected with the phase of the field $\phi$. 
This motivates us to parametrize the complex scalar field in 
terms of an amplitude and a phase as $\phi=\rho e^{i\varphi}/\sqrt{2}$, such that the Lagrangian of Eq. (\ref{eq:abh}) becomes, 
\begin{eqnarray}
\label{Eq:L-unitary-gauge}
\mathcal{L}_b&=&i\frac{\theta}{4\pi^2}\epsilon_{\mu\nu\lambda}a_\mu\partial_\mu a_\lambda+\frac{\rho^2}{2}(\partial_{\mu}\varphi-a_\mu)^2
\nonumber\\
&+&\frac{1}{2}(\partial_\mu\rho)^2+\frac{m_B^2}{2}\rho^2+\frac{\lambda}{8}\rho^4.
\end{eqnarray}
%
After performing the gauge transformation $a_\mu\to a_\mu+\partial_\mu\varphi$ and accounting for the periodic character of $\varphi$, 
we integrate out $a_\mu$ exactly to obtain, 
\begin{eqnarray}
\label{Eq:Seff-gauge-integrated}
S_{\rm eff}&=&\frac{1}{2}{\rm Tr}\ln\left(-i\frac{\theta}{2\pi^2}\epsilon_{\mu\nu\lambda}\partial_\lambda+\rho^2\delta_{\mu\nu}\right)
-\frac{1}{2}{\rm Tr}\ln \rho^2
\nonumber\\
&+&\frac{\theta^2}{4\pi^4}\int d^3x\int d ^3x'D_{\mu\nu}(x,x')V_\mu(x)V_\nu(x')
\nonumber\\
&+&\int d^3x\left[\frac{1}{2}(\partial_\mu\rho)^2+\frac{m_B^2}{2}\rho^2+\frac{\lambda}{8}\rho^4\right],
\end{eqnarray}
where $D_{\mu\nu}(x,x')$ is the inverse of the operator $-i\frac{\theta}{2\pi^2}\epsilon_{\mu\nu\lambda}\partial_\lambda+\rho^2\delta_{\mu\nu}$ and, 
\begin{equation}
V_\mu(x)=\epsilon_{\mu\nu\lambda}\partial_\nu\partial_\lambda\varphi(x)
=2\pi\sum_a n_a\oint dy_\mu^a\delta^3(x-y^a),
\end{equation}
is the vortex loop current, with $n_a$ being the vortex quantum. The term $(1/2){\rm Tr}\ln\rho^2$ arises from the Jacobian of the transformation 
from complex fields to $\phi=\rho e^{i\varphi}/\sqrt{2}$. It cancels out against explicit calculation of the first term of 
(\ref{Eq:Seff-gauge-integrated}) in the unitary gauge \cite{Nogueira-Kleinert-book}. 

For later use in the analysis of the duality using a lattice model, we integrate out the amplitude fluctuations approximately at one-loop order. This is easily 
done by considering the Gaussian fluctuations around around $\rho_0=2|\phi_{0,{\rm min}}|$ in the effective action (\ref{Eq:Seff-gauge-integrated}), i.e., we 
consider $\rho=\rho_0+\delta\rho$ integrate out the Gaussian fluctuations in $\delta\rho$. The result adds the following 
contribution to the effective action, 
\begin{equation}
\label{Eq:Gaussian-flucts}
\delta S_{\rm eff}=\frac{1}{2}{\rm Tr}\ln\left(-\Delta+m_B^2+\frac{3\lambda}{2}\rho_0^2\right). 
\end{equation}
A more accurate result would involve replacing $\rho_0^2$ in the above equation by $\langle\rho^2\rangle$ \cite{zinn2002quantum} and even more 
precise is to consider the full response and have the phase stiffness $\rho_s$ appearing as a coefficient of $(\partial_\mu\varphi-a_\mu)^2$.  
It is now instructive to recall a well known random path representation for 
the above result \cite{kleinert1989gauge,itzykson1991statistical,THOMAS1978513}. In this case we write the cutoff in terms of
 the shortest element of the path, $a$, which we later identify to the lattice spacing, so we can write $\Lambda=\pi/a$. 
 Denoting $P(\tilde{L})$ the number of closed paths of length $\tilde{L}$, we can use the known results of Refs. \cite{kleinert1989gauge,THOMAS1978513} 
 to write, 
 \begin{equation}
 \delta S_{\rm eff}=-\int_{0}^\infty dLP(\tilde{L})e^{-\epsilon \tilde{L}},
 \end{equation}
 where, 
 \begin{equation}
 \label{Eq:Line-energy}
 \epsilon=\frac{a}{6}\left(m_B^2+\frac{3\lambda}{2}\rho_0^2\right)+\frac{\ln 6}{a}. 
 \end{equation}

The particle-vortex duality will map the particle closed paths to vortex loops. 
We can use the well known general expression for the partition function for a statistical ensemble of vortex loops  
as derived from particles random worldlines \cite{kleinert1989gauge,THOMAS1978513}, 
\begin{equation}
\label{Eq:Z-random-walk}
Z=\sum_{N=0}^\infty\frac{1}{N!}\prod_{j=1}^N\sum_{\{C_j\}}e^{-S_{\rm vortex}-\epsilon \tilde{L}(C_j)},
\end{equation}
where $S_{\rm vortex}$ is the action yielding the (long-range) interaction energy between two loops $C_i$ and $C_j$, $\tilde{L}(C_j)$ is the length 
of the $j$-th loop and here $\epsilon$ is identified to the vortex core energy \cite{kleinert1989gauge,popov1973quantum}.

\section{Wilson fermions in three-dimensional Euclidean spacetime}\label{sec:wilsonfermions}

In this section we determine the partition function of a free massive Dirac theory in a three-dimensional cubic euclidean lattice. 
In order to put the Dirac theory in the lattice we use Wilson fermions \cite{Wilson1977}; see also Ref. \cite{Rothe2012}. 
This results in lattice action for the free massive Dirac fermions
\begin{align}
	&S=a^2\bigg[(m_0a+3R)\sum_n\bar{\psi}_n\psi_n {\notag}\\
	&-\frac{1}{2}\sum_{n\mu}[\bar{\psi}_n(R-\gamma_\mu)\psi_{n+\hat{\mu}} U_{\mu n}+\bar{\psi}_{n+\hat{\mu}}(R+\gamma_\mu)\psi_n U^\dagger_{\mu n-\hat{\mu}}]\bigg]	
	\label{eq:wilsonaction}
\end{align}
where $R$ is the Wilson parameter and $a$ is the lattice spacing, which we  set to unity until otherwise specified. The euclidean Dirac matrices 
above are given by the Pauli matrices. 

In the continuum limit \cref{eq:wilsonaction} will converge to (\ref{eq:fbduality-1}) independent of the value of the $R$ as long as $R\neq0$. 
We also assume that $m_0>0$. 
The coupling of the Wilson fermions to the field $U_{\mu n}=\exp(-iA_{\mu,n})$ 
enforces the local gauge invariance of Eq. (\ref{eq:wilsonaction}) in the lattice.  
Rewriting Eq. (\ref{eq:wilsonaction})  as \cite{Stamatescu1982}, we find
\begin{subequations}
	\begin{align}
		S&=\frac{1}{2\kappa}\sum_{nm}\bar{\psi}_nK_{nm}[U]\psi_m\\
		K[U]_{nm}&=\delta_{nm}\mathbbm{1}-\kappa M_{nm}[U]\\
		M_{nn+\hat{\mu}}[U]&=(R-\gamma_\mu) U_{\mu n}\label{eq:ma}\\
		M_{nn-\hat{\mu}}[U]&=(R+\gamma_\mu)U^\dagger_{\mu n-\hat{\mu}}\label{eq:mb},
	\end{align}
\end{subequations}
where $\kappa=\frac{1}{2(m_0+3R)}$ and \cref{eq:ma} and \cref{eq:mb} gives non-zero elements of $M$.
This leads to the lattice partition function
\begin{subequations}
	\begin{align}
		Z&=\int\mathcal{D}{\bar{\psi}\mathcal{D}\psi}e^{-\frac{1}{2\kappa}\sum_{nm}\bar{\psi}_nK_{nm}[U]\psi_m}\\
		&=\det\bigg[\frac{1}{2\kappa}K[U]\bigg]\\
		&=\exp\bigg[-\sum_{n=1}^\infty\frac{\kappa^{2n}}{2n}\text{Tr}\Big[M^{2n}[U]\Big]\bigg] \label{eq:serlatdet},
	\end{align}
\end{subequations}
where we have dropped a constant term $\exp[-\text{Tr}[\ln(2\kappa)]]$ 
in the last step. We only sum over even number of products of $M$'s, since the trace of an odd number $M$'s vanishes, as it 
is apparent from \cref{eq:ma} and \cref{eq:mb}.  The convergence of this series is explicitly discussed in \cref{sec:convfermiondet}. 

In the next steps we will calculate $\text{Tr}[M^{2n}]$. Being a trace, only paths forming closed loops contribute. Thus,  
\begin{equation}
Z=\exp\bigg[-\sum_{n=1}\sum_{\{C_{2n}\}}\frac{\kappa^{2n}}{I(C_{2n})}\text{tr}[M_{C_{2n}}]\bigg] \label{eq:pf}
\end{equation}
where the trace symbol ${\rm tr}[\dots]$ is over products of Pauli matrices and should not be confused 
with ${\rm Tr}[\dots]$, which denotes the trace over spacetime indices in the lattice. $M_{C_{2n}}$ is a $2n$-fold 
path-ordered product of  $M$'s along a  path $C_{2n}$ consisting of $2n$ number of lattice sites. From now on we choose the 
value of the Wilson parameter to be $R=1$. One main advantage of this choice is that now 
summation will be over all possible connected, non-backtracking paths of length $2n$ with $I(C_{2n})$ being the number of times a  
fermion travels along the path $C_{2n}$. Thus, $I(C_{2n})$ plays the role of a winding number. Simple examples of loops are illustrated in \cref{fig:loops}. 
Backtracking  paths like the one arising in the loop of  \cref{fig:loops}-d 
yield no contribution when $R=1$, since in this case $R\pm\gamma_\mu$ are projection 
operators and we have $(1-\gamma_\mu)(1+\gamma_\mu)=0$.  This special choice of $R$ is irrelevant in the scaling limit implied by 
the continuum model, which corresponds to the regime where $m_0a\ll 1$. 

From here on forward we take $A_{\mu n}=0$ and define the projector, 
\begin{equation}
\Gamma(\vec{e}(i))=1-\vec{\gamma}\cdot\vec{e}(i)
\end{equation}
where $\vec{e}(i)$ is a unit vector tangent to a $i$-th segment of a given path. Thus,  
\begin{equation}
\text{tr}[M_{C_{2n}}]=\text{tr}\bigg[\prod^{2n}_{i=1}\Gamma(\vec{e}(i))\bigg],
\end{equation}
yielding \cite{Stamatescu1982},
\begin{align}
	&\text{tr}\bigg[\prod^{2n}_{i=1}\Gamma(\vec{e}(i))\bigg]=\text{tr}\bigg[\prod^{k}_{i=1}\Gamma(\vec{e}(i))^{p_k}\bigg]\ \text{with}\  \vec{e}(i)\neq\vec{e}(i+1) {\notag}\\&=2^{2n-k}\text{tr}\bigg[\prod^{k}_{i=1}\Gamma(\vec{e}(i))\bigg]\ \text{with}\  \vec{e}(i)\neq\vec{e}(i+1),
\end{align} 
where $k$ are the number of straight sections (or sides) of a given loop,  $p_k$ is the length of the $k$-th straight section and in the last equality we used   $\Gamma(\vec{e}(i))^n=2^{n-1}\Gamma(\vec{e}(i))$. 

\begin{figure}
\includegraphics{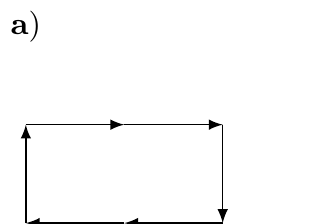}
\includegraphics{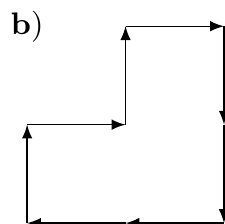}
\includegraphics{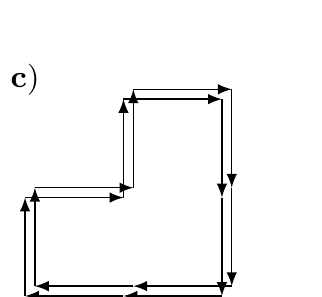}
\includegraphics{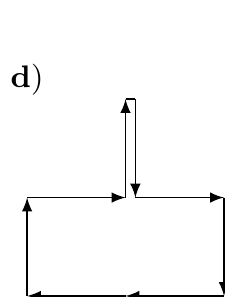}
	\caption{\textbf{a)} For this loop the perimeter $L$ is equal to the number of lattice sites $2n=6$, number of straight sections/sides is $k=4$, and number of windings $I=1$. \textbf{b)} Here, $2n=8$, $k=6$ and $I=1$ (number of winding would not change with orientation of the loop). \textbf{c)} Here, $2n=16$ and $I=2$ and $k=12$. \textbf{d)} Non-allowed loop, since it consists of backtracking. }\label{fig:loops}
\end{figure}

Next, we parametrize the unit tangent vectors in spherical coordinates as
\begin{equation}
\vec{e}(i)=(\sin\theta_i\cos\phi_i,\sin\theta_i\sin\phi_i,\cos\theta_i), \label{eq:runit}
\end{equation} 
and represent  the eigenstates of the 
vector of Pauli matrices as $|\vec{e}(i);\pm\rangle$, which satisfy,
\begin{equation}
\vec{\gamma}\cdot\vec{e}(i)|\vec{e}(i);\pm\rangle=\pm|\vec{e}(i);\pm\rangle,
\end{equation}
implying in this way, 
\begin{equation}
\Gamma(\vec{e}(i))=2|\vec{e}(i);-\rangle\langle\vec{e}(i);-|.
\end{equation}
Therefore, 
\begin{equation}
\text{tr}\bigg[\prod^{k}_{i=1}\Gamma(\vec{e}(i))\bigg]=2^k\prod^k_{i=1}\langle\vec{e}(i);-|\vec{e}(i+1);-\rangle \label{eq:tracefor}
\end{equation}
where  $|\vec{e}(k+1);-\rangle=|\vec{e}(1);-\rangle$. From now on, we denote $|\vec{e}(i);-\rangle\equiv|\vec{e}(i)\rangle$ for notational simplicity. The inner product can be written in terms of amplitude and phase, as
\begin{equation}
\langle\vec{e}(i)|\vec{e}(i+1)\rangle=|\langle\vec{e}(i)|\vec{e}(i+1)\rangle|\exp(i\arg[\langle\vec{e}(i)|\vec{e}(i+1)\rangle]).
\end{equation} 
Note that $|\langle\vec{e}(i)|\vec{e}(i+1)\rangle|=\frac{1}{\sqrt{2}}$ for all $i$, which can be seen by simply calculating the magnitude of 
the inner product between all distinct pairs. 
For the phase factor we have 
\begin{align}
	\arg\langle\vec{e}(i)|\vec{e}(i+1)\rangle&=\arctan\bigg[\frac{\sin\Delta\phi_i\cot\frac{\theta_i}{2}\cot\frac{\theta_{i+1}}{2}}{1+\cos{\Delta\phi_i}\cot\frac{\theta_i}{2}\cot\frac{\theta_{i+1}}{2}}\bigg]{\notag}\\
	&=\frac{\Omega_i'}{2}, \label{eq:sptri}
\end{align}where the $\Omega_i'$  is the area of an spherical triangle on a unit sphere\footnote{Even though we call that an area it can have a negative value.} \cite{todhunter1863spherical}, 
which is shown on \cref{fig:unitsphere} with the corners defined by the unit vectors $-\hat{z}$, $\vec{e}(i)$ and  $\vec{e}(i+1)$. Now if we combine \cref{eq:sptri} with  \cref{eq:tracefor} we obtain
\begin{align}
	\text{tr}\bigg[\prod^{k}_{i=1}\Gamma(\vec{e}(i))\bigg]&=2^{k/2}\prod^k_{i=1}\exp\bigg(i\frac{\Omega'_i}{2}\bigg){\notag}\\
	&=2^{k/2}\exp{i\frac{\Omega'}{2}}\label{eq:tracetosolidangle}
\end{align}
\begin{figure}
	\includegraphics{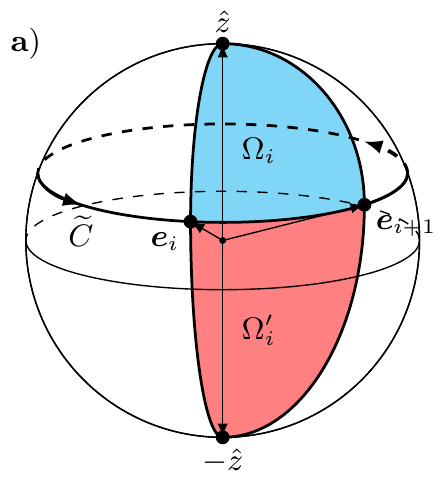}
	\hspace{0.3cm}
	\includegraphics{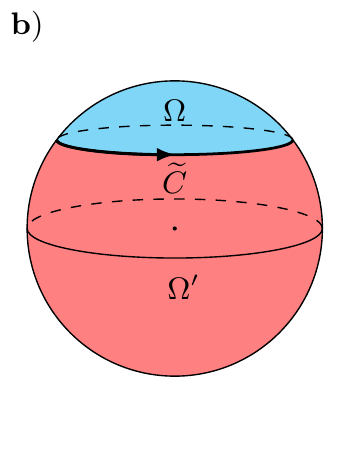}
	
	\caption{\textbf{a)} For a given $\widetilde{C}$,  $\Omega_i$ is the  solid angle traced  by $\vec{e}_i$ while travelling from $i$ to $i+1$ \textbf{b)} $\Omega'$ is $4\pi-\Omega$. }\label{fig:unitsphere}
\end{figure}
Note that $\vec{e}$ is a unit tangent to the path $C_{2n}$. Now if we consider a moving frame on path $C_{2n}$, we define $\widetilde{C}$ as the path such that 
the  tip of  $\vec{e}$ draws in this moving frame see \cref{fig:cvsct}.
\begin{figure}
	\includegraphics{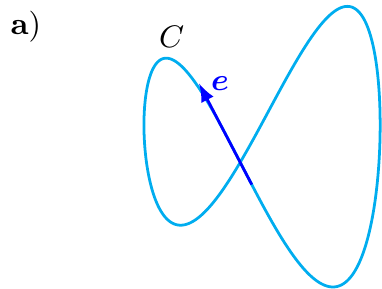}
	\hspace{0.3cm}
	\includegraphics{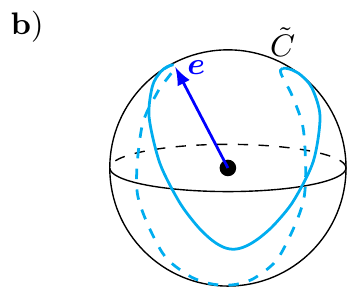}
	
	\caption{\textbf{a)} Here $C$ is a world line of a fermion and $\vec{e}$ is the unit tangent to $C$. \textbf{b)}  $\tilde{C}$ is the curve 
		drawn by the tip of $\vec{e}$ while travelling on $C$.}\label{fig:cvsct}\end{figure}
As it is shown in \cref{fig:unitsphere}, if we define $\Omega$ as the total solid angle acquired by $\vec{e}$ while travelling the given loop, then $\Omega'=4\pi-\Omega$. This can be better seen if we assume for a moment a continuous case such that $\Delta\phi_i\ll1$ and $\theta_i\sim\theta_{i+1}$. 
In this case we would have 
\begin{equation}
\Omega'_i\sim\Delta\phi_i(1+\cos\theta_i), \label{eq:crossec}
\end{equation}
and thus \begin{align}
	\exp{i\frac{\Omega'}{2}}&=\exp\bigg[{\frac{i}{2}\int_{\widetilde{C}} ds \dot{\phi}(1+\cos{\theta})}\bigg]{\notag},\\
	&=\exp(\frac{i}{2}[4\pi-\Omega]){\notag},\\
	&=\exp(-\frac{i}{2}\Omega). \label{eq:solidangle}
\end{align}
Thus, after Taylor expanding the exponential in \cref{eq:pf},  we can express  the partition function as, 
\begin{equation}
Z=\sum_{N=0}^{\infty}\frac{1}{N!}\prod_{j}^{N}\bigg[\sum_{\{C_j\}}-2^{L[C_j]-k_j/2}\frac{\kappa^{L[C_j]}}{I_{C_{j}}}e^{-\frac{i}{2}\Omega_{C_{j}}}\bigg] \label{eq: pfsolid}
\end{equation}
where $\sum_{\{C_j\}}$ is summation over all non-back tracing connected loops and 
$L[C_j]$ is the perimeter of the loop, which is simply given by 
the number of sites on the loop.  In \cref{eq: pfsolid} we sum over all possible loops, where  some of these  loops trace a path several times as it is shown in \cref{fig:loops}-c. We can rewrite \cref{eq: pfsolid} in an equivalent form where we sum  over loops with single winding as,
\begin{align}
	&Z={\notag}\\&\sum_{N=0}^{\infty}\frac{1}{N!}\prod_{j}^{N}\bigg[\sum_{\{C_j\}'} \sum_{n_j=1}^{\infty}-2^{n_jL[C_j]-\frac{n_jk_j}{2}}\frac{\kappa^{n_jL[C_j]}}{n_j}e^{-\frac{i}{2}n_j\Omega_{C_{j}}}\bigg] \label{eq: pfsolid2}
\end{align}
where $\sum_{\{C_j\}'}$ denotes the sum over loops with single windings and $n_j$ is the winding number. 

Finally, we can write the solid angle $\Omega$ swept out by $\vec{e}$ in terms of writhe of the curve. 
To do so we follow the approach of Refs. \cite{Hashimoto1989,Frank-Kamenetskii1981}. Assume that we have a closed curve $C_1$ of length $L$ with parametrization $\vec{r}_1(s)$ where $0\leq s\leq L$. We also set $s$ as the arc length between $\vec{r}_1(0)$ and $\vec{r}_1(s)$ of the curve, which implies $|\dot{\vec{r}}_1|=1$. Now assume that we have another curve $C_2$, with parametrization  $\vec{r}_2$, such that $\vec{r}_2=\vec{r}_1+\epsilon\vec{a}$, where $\vec{a}$ is the unit vector normal to $\vec{r}_1$ i.e. $\vec{r}_1\cdot\vec{a}=0$, and $\epsilon$ is an infinitesimal constant; see Fig. \ref{fig:twist}.
\begin{figure}
	\includegraphics{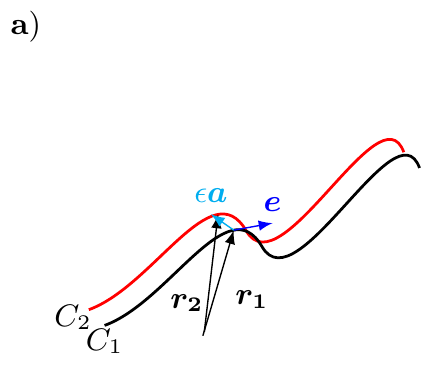}
	\includegraphics{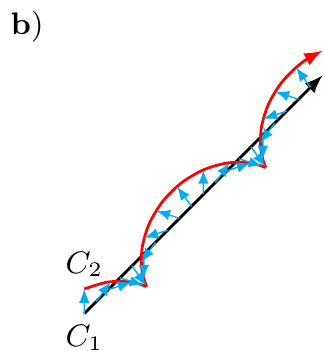}
	\caption{\textbf{a)} Here $\vec{e}$ is the tangent vector at point $\vec{r}_1$ on curve $C_1$ and  $\epsilon\vec{a}$ is the distance between $C_2$ and $C_1$,  where $\vec{a}\cdot\vec{r}_1=0$. \textbf{b)} If $C_2$ winds around $C_1$, $\vec{a}$ rotates around $C_1$. If we view $C_1$ and $C_2$ as edges of a ribbon, then $\vec{a}$ rotates as the ribbon twists. The expression  $\frac{\vec{a}\times\vec{\dot{a}}\epsilon^2}{\epsilon^2}\cdot{\vec{e}}$ can be regarded as the angular speed of the point $\epsilon\vec{a}$ around $\vec{e}$, and the expression for the twist as  given in \cref{eq:twist} is the angular displacement of $\epsilon\vec{a}$ divided by $2\pi$.\label{fig:twist}}
\end{figure}
Now recall the  
C{\u{a}}lug{\u{a}}reanu-White theorem \cite{calugareanu1959integrale,calugareanu1961classes,White},  
which relates and defines the linking number $G$, the writhe $\mathcal{W}$, and the twist $T$ of two curves as \cite{Hashimoto1989,Frank-Kamenetskii1981},
\begin{subequations}
	\begin{align}
		G[C_1,C_2]&=\mathcal{W}[C_1]+T[C_1,C_2],\label{eq:ct}\\
		G[C_1,C_2]&=\frac{1}{4\pi}\oint_{C_1}\oint_{C_2}\frac{d\vec{r}_1\times d\vec{r}_2\cdot[\vec{r}_1-\vec{r}_2]}{|\vec{r}_1-\vec{r}_2|^3},\\
		\mathcal{W}[C_1]&=\frac{1}{4\pi}\oint_{C_1}\oint_{C_1}\frac{d\vec{r}_1\times d\vec{r}_2\cdot[\vec{r}_1-\vec{r}_2]}{|\vec{r}_1-\vec{r}_2|^3},\\
		T[C_1,C_2]&=\frac{1}{2\pi}\int_{0}^{L} ds [\vec{{a}}({s})\times\vec{\dot{a}}({s})]\cdot\vec{\dot{r}}_1({s}).\label{eq:twist}
	\end{align}	\label{eq:cwtheo}
\end{subequations}

We now seek to relate the writhe $\mathcal{W}$ to $\Omega$ and proceed by first relating the twist $T$ to $\Omega$ and then use Eq.~(\ref{eq:ct}). Since $|\dot{\vec{r}}_1|=1$, $\dot{\vec{r}}_1$ is the unit tangent vector to curve $C_1$, i.e. $\dot{\vec{r}}_1(s)=\vec{e}(s)$, and we parametrize it as $\vec{e}(s)=\hat{\vec{e}}_r(s)$, where $\hat{\vec{e}}_r(s)$ is a radial unit vector in spherical coordinates as in \cref{eq:runit}, and we can choose the frame vector \cite{Hashimoto1989} $\vec{a}(s)=\hat{\vec{e}}_\phi(s)$ where $\hat{\vec{e}}_\phi(s)$ is the azimuthal unit vector in spherical coordinates. We then evaluate \cref{eq:twist} as,
\begin{align}
	T[C_1,C_2]&=\frac{1}{2\pi}\oint_{C_1}d\vec{r}_1\cdot[\vec{{a}}({s})\times\vec{\dot{a}}({s})]{\notag}\\
	&=\frac{1}{2\pi}\int_{0}^{L} ds \dot{\phi}(s)\cos{\theta(s)}.
\end{align}
Next, the solid angle traced by $\hat{\vec{e}}_r$ while traveling on $C_1$ is given as,
\begin{subequations}
	\begin{align}
		\Omega&=\oint_{\widetilde{C}_1}d\vec{e}_r\cdot\hat{\vec{e}}_{\phi}\frac{1-\cos\theta}{|e_r|\sin\theta},\label{eq:bolbol1}\\
		&=\int_0^Lds\dot{\phi}(s)(1-\cos{\theta}(s)),\label{eq:bolbol2}\\
		&=2\pi u-\int_{0}^{L} ds \dot{\phi}(s)\cos{\theta(s)},\quad u\in\mathbb{Z},\label{eq:bolbol3}\\
		&=-2\pi T[C_1,C_2] \mod(2\pi).\label{eq:bolbol4}
	\end{align}
\end{subequations}
Note that in \cref{eq:bolbol1} we take the line integral over a vector potential of a Dirac monopole along the curve  $\widetilde{C}_1$, which yields the solid angle  
associated to  
$\widetilde{C}_1$, which corresponds to the solid angle swept by $\vec{e}$ traveling along $C_1$.   
In \cref{eq:bolbol3} we used the fact that $\vec{r}_1(0)=\vec{r}_1(L)$. Finally, we substitute \cref{eq:bolbol4} to \cref{eq:ct} to get,
%
\begin{equation}
\mathcal{W}[C]=\frac{\Omega}{2\pi}+p,\label{eq:writhetomega}
\end{equation} 
where $p=G-u$. Finally, we can use the observation that $p$ is an odd integer \cite{Hashimoto1989}. 

Now by putting \cref{eq:writhetomega} into \cref{eq: pfsolid2} we obtain the exact fermionic partition function as a series in $\kappa$\ of the form
\begin{subequations}
	\begin{align}
		Z&=\sum_{N=0}^{\infty}\frac{1}{N!}\prod_{j}^{N}\bigg[\sum_{\{C_j\}'}\sum_{n_j=1}^{\infty}z[C_j,n_j]\bigg]\label{eq:et}\\
		z&=-2^{n_jL[C_j]-\frac{n_jk_j}{2}}\frac{\kappa^{n_jL[C_j]}}{n_j}(-1)^{n_j}e^{-i\pi n_j\mathcal{W}[C_{j}]},
	\end{align}
\end{subequations}
where $C_j$ denotes a closed loop with straight sections $k_j$,  length $L[C_j]$, winding number $n_j$ and, most importantly, writhe $\mathcal{W}[C_j]$. 

As a side note, the above result is closely related to the usual loop representation of the fermionic determinant in terms of loops, 
within a so called the hopping parameter expansion (more details can be found in section 5.1.3 of Ref. \cite{montvay}). 

\section{Convergence of the Lattice Fermionic Determinant}
\label{sec:convfermiondet}

The series expansion for the lattice determinant for Wilson fermions above in \cref{eq:serlatdet} raises the question as to its radius of convergence, and in particular the critical value of $\kappa$. For this discussion we adopt the approach of Ref. \cite{Gattringer1999}. Since the term in the exponential of  \cref{eq:serlatdet} is the series expansion of $\text{Tr}\big[\ln\big[1-\kappa M\big]\big]$, the series converges for 
\begin{equation}
\kappa||M||_\infty<1,
\label{eq:radofcon}\end{equation}
where the infinity norm is given as the square root of the largest \footnote{The absolute value of the eigenvalue is largest.} eigenvalue of the $M^\dagger M$ \cite{Gattringer1999}. In order to find it, we first use the Fourier representation of $M$,
\begin{align}
	M_{nm}&=\sum_{\mu}\int_{-\pi}^{\pi}\frac{d^3p}{(2\pi)^3}e^{ip(n-m)}2[\cos(p_\mu)-i\gamma_\mu\sin(p_\mu)]{\notag}\\
	&=\int_{-\pi}^{\pi}\frac{d^3p}{(2\pi)^3}e^{ip(n-m)}\sum_{\alpha=\pm}|\alpha\rangle\langle\alpha|E_\alpha(p)
\end{align}
where in the last line we write $M$ in a diagonal form  using the  eigenvectors $|\alpha\rangle$  of $M_{nm}$, and $E_\pm(p)=2\big[\sum_{\mu}\cos(p_\mu)\pm i[\sum_{\mu}\sin^2(p_\mu)]^{1/2}\big]$. 
Thus, we can Fourier transform $M^{\dagger}M$ as
\begin{align}
	(M^\dagger M)_{nm}=\int_{-\pi}^{\pi}\frac{d^3p}{(2\pi)^3}e^{ip(n-m)}\sum_{\alpha}|\alpha\rangle\langle\alpha||E_\alpha(p)|^{2}
\end{align}
where $|E(p)|^2=4\big[[\sum_{\mu}\cos(p_\mu)]^2+ \sum_{\mu}\sin^2(p_\mu)\big]$. Thus  $|E(0)|^2=36$ is the highest eigenvalue and the norm is 
$
||M||_\infty=6.
$
We parametrize $\kappa$ as 
\begin{equation}
\kappa=\frac{1}{6}e^{-m}\label{eq:lambdapat}
\end{equation} where $m=\ln(1+\frac{m_0}{3})$ then using \cref{eq:radofcon} we find that the series converges for 
$m_0>0$. 
Thus not surprisingly, we find that the series converges only for a nonzero bare mass. We can now rewrite \cref{eq:et} using \cref{eq:lambdapat} as,
\begin{subequations}
	\begin{align}
		Z&=\sum_{N=0}^{\infty}\frac{1}{N!}\prod_j^N\bigg(\sum_{\{C_j\}'}\sum_{n_j=1}^{\infty}z[C_j,n_j]\bigg).\\
		z&=-\frac{2^{L[C_j]n_j-n_jk_j[C_j]/2}}{6^{n_jL[C_j]}}\frac{(-1)^{n_j}}{n_j}e^{-mn_jL[C_j]} e^{-i\pi n_j\mathcal{W}[C_{j}]}.
	\end{align}\label{eq:fermioniccase}
\end{subequations}
This final form of the fermionic partition sum expressed in terms of the writhe is our first key result.

\section{Particle-vortex duality in the Chern-Simons lattice Abelian Higgs model}
\label{sec:pv-duality}
Having studied the bosonic continuum actions of Eqs.~\eqref{Eq:pvduality} in Sec.~\ref{Sec:dualbos}, and given form of the partition sum of the fermionic action in terms of the writhe derived in the last section, we aim to establish the bosonization duality between the lattice versions of Eqs.~\eqref{Eq:pvduality} and \eqref{eq:fbduality-1} by also expressing the bosonic partition sum in terms of the writhe. We proceed in three steps: first, we connect the bosonic particle-vortex duality in the continuum to its lattice equivalent, then calculate the bosonic partition sum in the dual bosonic action, and finally compare it with the fermionic result.

\subsection{Particle-vortex duality on the lattice}\label{sec:PVD}

The partition function for the Abelian CS Higgs model on the lattice with a non-compact gauge field is given by, 
\begin{equation}
\label{Eq:Partition-cosine}
Z=\left[\prod_{j,\mu}\int_{0}^{2\pi}\frac{d\varphi_j}{2\pi}\int_{-\infty}^{\infty} da_{j\mu}\right]\sum_{\{n_{j\mu}\}}e^{-S},
\end{equation}
where the lattice action is given in the Villain approximation as, 
\begin{eqnarray}
\label{Eq:Action-cosine}
S&=&\sum_{j}\left[i\frac{\theta}{4\pi^2}\epsilon_{\mu\nu\lambda}a_{j\mu}\Delta_\nu a_{j\nu}
\right. \nonumber\\
&+&\left.\frac{J}{2}(\Delta_\mu\varphi_j-2\pi n_{j\mu}-a_{j\mu})^2\right],
\nonumber\\
\end{eqnarray}
where $\sum_{\{n_{i\mu}\}}=\prod_{i\mu}\sum_{n_{i\mu}=-\infty}^\infty$, 
$J>0$ is the bare phase stiffness and $\Delta_\mu$ represents the forward discrete derivative, $\Delta_\mu f_i=f_{i+1}-f_i$. 
In the above action the integer valued lattice fields $n_{j\mu}$ enforces the periodicity of $\varphi_j$ fulfilling the integer (or vortex) gauge 
invariance, $\varphi_j\to\varphi_j+2\pi L_{j\mu}$, $n_{j\mu}\to n_{j\mu}+L_{j\mu}$, where $L_{j\mu}$ is an arbitrary integer. 
This discrete gauge invariance is a common feature of the so called Villain action \cite{kleinert1989gauge} and 
arises here in addition to the usual gauge invariance associated to the lattice gauge field $a_{j\mu}$.

In order to derive the dual model, we follow closely the approach of Ref. \cite{Peskin1978} and use the Poisson summation formula, 
\begin{eqnarray}
\label{Eq:Poisson}
&&(2\pi a)^{1/2}\sum_{n=-\infty}^{\infty} 
e^{-\frac{a}{2}(x-2\pi n)^2}
\nonumber\\
&=&\sum_{m=-\infty}^{\infty}e ^{-\frac{1}{2a}m^2+ixm},
\end{eqnarray}
to introduce an auxiliary integer-valued lattice field $J_{j\mu}$, i.e., 
\begin{eqnarray}
&&\sum_{\{n_{j\mu}\}}e^{-\frac{J}{2}(\Delta_\mu\varphi_j-2\pi n_{j\mu}-a_{j\mu})^2}
\nonumber\\
&\sim&\sum_{\{J_{j\mu}\}}e^{-\frac{1}{2J}J_{j\mu}^2+iJ_{j\mu}(a_{j\mu}-\Delta_\mu\varphi_j)}.
\end{eqnarray}
Next we use summation by parts in the term  $-J_{j\mu}\Delta_\mu\varphi_j$ to convert it to 
$\Delta_\mu J_{j\mu}\varphi_j$, which allows us to 
integrate the phase variables out 
to obtain the zero divergence constraint, $\Delta_\mu J_{j\mu}=0$. The latter implies that we are 
dealing with a sum over configurations where the vortices form loops \cite{kleinert1989gauge}.  The action  is thus rewritten as 
\begin{equation}
\label{Eq:Action-1}
S'=\sum_{j}\left[i\frac{\theta}{4\pi^2}\epsilon_{\mu\nu\lambda}a_{j\mu}\Delta_\nu a_{j\nu}+\frac{1}{2J}J_{j\mu}^2
-iJ_{j\mu} a_{j\mu}.
\right].
\end{equation}
The constraint is solved by introducing the curl of another 
integer field, $M_{j\mu}$, such that $J_{j\mu}=\epsilon_{\mu\nu\lambda}\Delta_\nu M_{j\lambda}$, which leads to, 
\begin{eqnarray}
\label{Eq:Action-2}
S''&=&\sum_{j}\left[i\frac{\theta}{4\pi^2}\epsilon_{\mu\nu\lambda}a_{j\mu}\Delta_\nu a_{j\nu}
+\frac{1}{2J}(\epsilon_{\mu\nu\lambda}\Delta_\nu M_{j\lambda})^2
\right.\nonumber\\
&-&\left.i(\epsilon_{\mu\nu\lambda}\Delta_\nu M_{j\lambda}) a_{j\mu}\right].
\end{eqnarray}
Thus, upon integrating out the gauge field $a_{j\mu}$, we obtain, 
\begin{eqnarray}
\label{Eq:Dual-Action-0}
Z&=&\sum_{\{M_{j\mu}\}}e^{-S'''}
\nonumber\\
S'''&=&\frac{1}{2}\sum_{j}\left[\frac{1}{J}(\epsilon_{\mu\nu\lambda}\Delta_\nu M_{j\lambda})^2
-i\frac{2\pi^2}{\theta}\epsilon_{\mu\nu\lambda}M_{j\mu}\Delta_\nu M_{j\lambda}\right].
\nonumber\\
\end{eqnarray}
We now use the Poisson summation formula in the form, 
\begin{equation}
\sum_{n=-\infty}^{\infty}f(n)=\sum_{m=-\infty}^\infty\int_{-\infty}^{\infty}\frac{dk}{2\pi}e^{i2\pi km}f(k),
\end{equation}
to convert the 
integer field $M_{j\mu}$ 
into a real-valued gauge field $b_{j\mu}$, and noting that this last step introduces another integer field $\widetilde{J}_{j\mu}$, we obtain finally 
the dual action in the form
\begin{eqnarray}
\label{Eq:Dual-Action}
\widetilde{S}&=&\frac{1}{2}\sum_{j}\left[\frac{1}{J}(\epsilon_{\mu\nu\lambda}\Delta_\nu b_{j\lambda})^2
-i\frac{2\pi^2}{\theta}\epsilon_{\mu\nu\lambda}b_{j\mu}\Delta_\nu b_{j\lambda}\right]
\nonumber\\
&-&i2\pi \sum_{j} \widetilde{J}_{j\mu}b_{j\mu},
\end{eqnarray}
after integrating out $a_{j\mu}$. Note that unlike the original action (\ref{Eq:Action-cosine}), the dual action above features a Maxwell term with the bare phase 
stiffness of the original model appearing as a gauge coupling. As a consequence of gauge invariance, the lattice vortex current 
field $\widetilde{J}_{j\mu}$ also has a vanishing divergence. 

By letting $J\to\infty$ in Eq.  (\ref{Eq:Dual-Action}), we see that for $\theta=\pi$ and 
after rescaling $b_{j\mu}\to b_{j\mu}/(2\pi)$ the latter is the same as the one arising in the 
partition function (\ref{Eq:Action-1}) up to the sign of the CS term. As far as the partition function is concerned, the 
sign of the CS term is immaterial, since integrating out 
either $a_{j\mu}$ or $b_{j\mu}$ in the $J\to\infty$ limit leads to the same vortex current interaction when $\theta=\pi$ as we sum over all 
integer-valued vortex currents. The theory is thus self-dual in this regime. 

If we  smear the constraint $\Delta_\mu\widetilde{J}_{j\mu}=0$ by adding the term $(\gamma/2)\widetilde{J}_{j\mu}^2$ to the dual 
action (\ref{Eq:Dual-Action}), corresponding  
to adding a chemical potential for the vortex loops \cite{Peskin1978,Dasgupta-Halperin_PhysRevLett.47.1556}, 
and apply once more the Poisson formula (\ref{Eq:Poisson}), 
we obtain, 
\begin{eqnarray}
\label{Eq:Dual-Action-1}
\widetilde{S}'&=&\frac{1}{2}\sum_{j}\left[\frac{1}{J}(\epsilon_{\mu\nu\lambda}\Delta_\nu b_{j\lambda})^2
-i\frac{2\pi^2}{\theta}\epsilon_{\mu\nu\lambda}b_{j\mu}\Delta_\nu b_{j\lambda}\right.
\nonumber\\
&+&\left.\frac{1}{\gamma}(\Delta_\mu\widetilde{\varphi}_j-2\pi N_{j\mu}-2\pi b_{j\mu})^2\right],
\end{eqnarray}
where $N_{j\mu}$ is an integer lattice field and $\widetilde{\varphi}_j$ is a phase variable originating from the integral 
representation of the Kronecker delta constraint enforcing $\Delta_\mu\widetilde{J}_{j\mu}=0$, i.e., 
\begin{equation}
\delta_{\Delta_\mu\widetilde{J}_{j\mu},0}=\int_{-\pi}^\pi\frac{d\widetilde{\varphi}_j}{2\pi}e^{i\widetilde{\varphi}_j\Delta_\mu\widetilde{J}_{j\mu}}. 
\end{equation}

Equation \eqref{Eq:L-unitary-gauge}, equation (\ref{Eq:Action-cosine}) and its dual form in Eq. (\ref{Eq:Dual-Action-1}) establishes a lattice version of the field theory duality of 
Eq. (\ref{Eq:pvduality}) in the regime where amplitude fluctuations of the bosonic fields are neglibile. The continuum limit of the lattice duality is expected to approach the field theory duality in the vicinity of the 
critical point.  However, we should emphasize that precise statements to this effect can only be achieved within the lattice formalism. 

\subsection{Duality and writhe}

In the case of the partition function for Wilson fermions we have seen that in order to make the writhe more apparent we had to partially 
evoke a continuum limit, while still counting fermionic loops configurations. We will employ a similar strategy for the dual bosonic 
partition function (\ref{Eq:Dual-Action}) in order to  
express it terms of linking and writhe numbers. Thus, the  continuum version of (\ref{Eq:Dual-Action}) can be obtained by  first writing a    
functional integral for a given configuration featuring $N$ vortex loops,  
\begin{subequations}
	\begin{align}
	Z&\propto\sum_{\{\widetilde{J}_{x\mu}\}}{}^{'}\int\mathcal{D}be^{-S_{\text{MCS}}}
	\\
	\label{Eq:MCS}
	S_{\rm MCS}&=\int d^3x\left[\frac{1}{2J}(\epsilon_{\mu\nu\lambda}\partial_\nu b_\lambda)^2
	\right.\nonumber\\
	&-\left.\frac{i\pi^2}{\theta}\epsilon_{\mu\nu\lambda} 
	b_\mu \partial_\nu b_\lambda-i2\pi  \widetilde{J}_{\mu} b_\mu\right],
	\end{align}
\end{subequations}
with, 
\begin{equation}
\label{Eq:Vortex-Loop}
\widetilde{J}_{\mu}(x)=\sum_{a=1}^{N}n_a\int_{0}^{\tilde{L}_a}ds \frac{dy_\mu^{a}(s)}{ds}\delta^3(x-y^{a}(s)), 
\end{equation}
where, $n_a\in\mathbb{Z}^+$ \footnote{We do not need to add negative values of $n_a$, since we can control the sign of vorticity with the orientation of the vortex loop.}  is the quantum number of the $a$-th vortex loop 
and $y_\mu^{a}(s)$, $s\in[0,\tilde{L}_a]$  is a parametrization of  the curve describing a loop $C_{a}$ with length $\tilde{L}_a$
satisfying the boundary conditions $y_\mu^{a}(0)=y_\mu^{a}(\tilde{L}_a)$. Equation (\ref{Eq:Vortex-Loop}) clearly satisfies $\partial_\mu \widetilde{J}_\mu=0$. 

We can write the partition function as a summation over all possible vortex current fields $\widetilde{J}$ configurations as;
\begin{equation}
Z=\sum_{N=0}^{\infty}\frac{1}{N!}\prod_{j=1}^{N}\sum_{\{C_j\}'}\sum_{n_j}Z(C_1,n_1;\dots;C_{N},n_{N}) \label{eq:allvorsum}
\end{equation}
where $Z(C_1,n_1;\dots;C_{N},n_{N})$ is a partition function with a fixed vortex configuration which consists of $N$ number of vortex loops, 
where the $a$-th loop has a shape $C_a$ and  vortex quantum number  $n_a$.  
Multiplication by $\frac{1}{N!}$ provides the symmetry factor preventing the overcounting of 
identical configurations, and $\sum_{\{C\}'}$ is summation over all connected non-backtracing loops with single winding and with both orientations. Explicitly we can write the partition function for a fixed configuration  as, 
\begin{equation}
Z(C_1,n_1;\dots;C_{N},n_{N})=\int\mathcal{D}b_\mu e^{-S_{\rm MCS}},
\end{equation}
normalized such that $Z(0)=1$. 
We now integrate out $b_\mu$, in momentum space so that
\begin{eqnarray}
\label{Eq:Z-JJ}
&&Z(C_1,n_1;\dots;C_{N},n_{N})=
\nonumber\\
&&\exp\left[-2\pi^2\int \frac{d^3p}{(2\pi)^3} D_{\mu\nu}(p)\widetilde{J}_\mu(p)\widetilde{J}_\nu(-p)\right],
\end{eqnarray}
where,
\begin{eqnarray}
D_{\mu\nu}(p)&=&\frac{J}{p^2+4\pi^4J^2/\theta^2}\left(\delta_{\mu\nu}-\frac{p_\mu p_\nu}{p^2}
\right.\nonumber\\
&-&\left.\frac{2\pi^2J}{\theta}\frac{\epsilon_{\mu\nu\lambda}p_\lambda}{p^2}
\right),
\end{eqnarray}
in the Landau gauge. Equation (\ref{Eq:Z-JJ}) give a vortex interaction identical to the one in Eq. (\ref{Eq:Seff-gauge-integrated}) when $\rho$ is 
uniform. 

In the limit where $J\to\infty$, which corresponds to $\Lambda\to\infty$ in the case ot $\rho_0$, we obtain
\begin{eqnarray}
\label{Eq:Z}
&&Z(C_1,n_1;\dots;C_{N},n_{N})=
\nonumber\\
&&\exp\left[\frac{i\theta}{4\pi}\sum_{a,b}n_an_b\oint_{C_a}\oint_{C_b}dy_\mu^{a} d y_\nu^{b}
\frac{\epsilon_{\mu\nu\lambda}(y_\lambda^{a}-y_\lambda^{b})}{|y^{a}- y^{b}|^3}
\right],
\end{eqnarray}
which in view of \cref{eq:cwtheo} can be rewritten as
 \begin{align}
 	\label{Eq:Z-W+G}
 	&Z(\{C\})={\notag}\\&\exp\left(i\theta \sum_an_a^2\mathcal{W}[C_a]+i2\theta\sum_{a<b}n_an_b G_{ab}\right),
 \end{align}
where we have introduced the notation, $\{C\}=\{C_1,n_1;\dots;C_{N},n_{N}\}$. 
For $\theta=\pi$ the term proportional to $G_{ab}$ does not contribute to the partition function due to the Gauss linking number theorem. 
If we use a statistical mechanical language and interpret $J$ as the exchange energy divided by the temperature and refer to the original 
lattice action (\ref{Eq:Action-cosine}), we see that the limit $J\to\infty$ corresponds to a zero temperature limit in this context. Furthermore, 
$J$ is related to the amplitude of the scalar field $\phi$, such that $J\sim \rho_0^2$, and we have seen in \cref{Sec:dualbos} that $\rho_0^2$ 
is indeed very large. 
Peskin \cite{Peskin1978} in his analysis of the particle-vortex duality referred to this regime as a "frozen superconductor".  
The analysis in Ref. \cite{Peskin1978} ignores the vortex core energy and adds it by hand as a small chemical potential for 
the vortices \cite{Peskin1978,Dasgupta-Halperin_PhysRevLett.47.1556}. However, the vortex core energy arises quite naturally, since it is related to 
the correlation length in the continuum theory. 
Furthermore, there is a direct relation between it and the bare mass, as we discussed in 
\cref{Sec:dualbos}. Therefore, the actual result corresponding to large $J$ is given by
\begin{align}
\label{Eq:Z-1}
&Z(\{C\})={\notag}\\&\exp\left(i\theta \sum_an_a^2\mathcal{W}[C_a]+i2\theta\sum_{a<b}n_an_b G_{ab}-\epsilon\sum_an_a^2\tilde{L}_a\right). 
\end{align}
where $\epsilon$ 
is the vortex core energy given by Eq. (\ref{Eq:Line-energy}).    
As in Eq. \eqref{Eq:Z-W+G}, for $\theta=\pi$ the second term does not contribute, since $G_{ab}\in\mathbb{Z}$ by 
virtue of the Gauss linking number theorem. 
Thus, after summing over all loop configurations we obtain the partition function as
\begin{equation}
Z=\sum_{N=0}^{\infty}\frac{1}{N!}\prod_{j=1}^{N}\sum_{\{C_j\}}\sum_{n_j}
\bigg[\exp(i\pi \mathcal{W}[C_j])\exp(-\epsilon \tilde{L}[C_j])\bigg]^{n_j^2}, \label{eq:bosoniccase}
\end{equation}
an equation that has the same form as Eq. (\ref{Eq:Z-random-walk}), with $S_{\rm vortex}=-i\pi\mathcal{W}$. The factor $\exp(i\pi \mathcal{W}[C_j])$ 
is crucial for the fermion-boson transmutation in 2+1 dimensions and is referred to as a  "spin factor" in the literature 
\cite{AMBJORN1990509,Grundberg1990,Goldman2018}.

\section{Comparison of fermionic and bosonic partition functions}
\label{sec:comparison}

Let us now compare the fermionic partition function with the bosonic one in the continuum. 
%
%
In order to take the continuum limit of lattice fermionic partition function, we first return to \cref{eq:tracefor} and introduce the  lattice spacing explicitly,
\begin{align}
\text{tr}[M_{C_{2n}}]&=\text{tr}\bigg[\prod^{2n}_{j=1}\Gamma(\vec{e}(ja))\bigg]=\prod^{2n}_{j=1}2\langle\vec{e}(ja)|\vec{e}((j+1)a)\rangle{\notag}\\
&=\prod^{2n}_{j=1}2\bigg[1+a\langle\vec{e}(s)|\frac{d}{ds}\vec{e}(s)\rangle\big|_{s=ja}+\mathcal{O}(a^2)\bigg]{\notag}\\
&=\prod^{2n}_{j=1}2\bigg[1+\frac{i}{2}a\dot{\phi}(s)(1+\cos(\theta(s)))+\mathcal{O}(a^2)\bigg]{\notag}\\
&\sim2^{\frac{\tilde{L}}{a}}\exp{-\frac{i}{2}\Omega[C_{\tilde{L}}]}
\end{align}  
where $\tilde{L}[C]=aL[C]$ and in the last line we used \cref{eq:solidangle}. In this case \cref{eq:fermioniccase} becomes, 
\begin{subequations}
	\begin{align}
	Z&=\sum_{N=0}^{\infty}\frac{1}{N!}\prod_{j}^{N}\bigg[\sum_{\{C_j\}'}\sum_{n_j=1}^{\infty}z[C_j,n_j]\bigg]\\
	z&=-{2^{\widetilde{L}[C_j]n_j/a}}e^{-i\pi n_jW[C_j]-(m+\ln 6)\widetilde{L}[C_j]n_j/a}\frac{(-1)^{n_j}}{n_j}.
	\end{align}
\end{subequations} 
We confine ourselves to the low-energy sector corresponding to $n_j=1$, 
in which case the fermionic partition function becomes,
\begin{equation}
Z_F=\sum_{N=0}^\infty\frac{1}{N!}\prod_{j=1}^N\sum_{\{C_j\}}{2^{\widetilde{L}[C_j]/a}}e^{-i\pi W[C_j]-(m+\ln 6)\widetilde{L}[C_j]/a}.
\end{equation}
 In the limit $m_0a\ll 1$, we have $m=\ln(1+m_0a/3)\approx m_0a/3$, which implies,  
\begin{equation}
\label{Eq:Z_F-cont}
Z_F\propto\sum_{N=0}^\infty\frac{1}{N!}\prod_{j=1}^N\sum_{\{C_j\}}e^{-i\pi W(C_j)-\widetilde M_0\tilde{L}(C_j)},
\end{equation}
where
\begin{equation}
\widetilde{M}_0=\frac{m_0}{3}+\frac{1}{a}\ln 3. 
\end{equation} 
The key observation is that the bosonic partition function \cref{eq:bosoniccase} \footnote{note that the difference of the overall sign in the writhe is immaterial, since we sum over loops with every possible orientations, 
with winding number $n_j=1$ precisely matches the fermionic one $Z_F$ in continuum limit when}
\begin{equation}
\epsilon=\widetilde{M}_0=\frac{m_0}{3}+\frac{1}{a}\ln 3.
\end{equation}
It is possible to connect this result to the continuum theory via the ensemble of paths discussed in Section \cref{Sec:dualbos}. 
Accordingly, 
by using Eq. (\ref{Eq:Line-energy}), we obtain a relation between the fermion, the boson mass $m_B$ and the $|\phi|^4$ coupling 
$\lambda$ of the continuum model, 
\begin{equation}
\label{Eq:mB2-m0}
\frac{m_0a}{3}=\frac{a^2}{6}\left(m_B^2+\frac{3\lambda}{2}\rho_0^2\right)+\ln 2. 
\end{equation}

Thus, we have  found that in the regime where the vortex energy entering the Boltzmann weight 
is minimum, 
corresponding to the winding $n_j=1$ in \cref{eq:bosoniccase} , 
the latter is identical to the fermionic partition function where in the fermion loop worldlines 
the particle travels the loop only once.  This establishes a correspondence between the fermionic particle worldline loops and vortex 
loops in the bosonic dual CS theory, which in turn sheds new light on  Polyakov's result for a first-quantized 
path integral description of massive Dirac fermions in 2+1 dimensions \cite{Polyakov_doi:10.1142/S0217732388000398}.

\section{Discussion}

The duality transformation from the Villain action (\ref{Eq:Action-cosine}) to (\ref{Eq:Dual-Action-1})  identifies in the continuum limit to the 
duality (\ref{Eq:pvduality}). In contrast to the situation in 3+1 dimensions, the Maxwell term is IR irrelevant in 2+1 dimensions, and 
therefore both theories in (\ref{Eq:pvduality}) flow to the same gauged (IR stable) Wilson-Fisher fixed point. This situation corresponds in the 
lattice to $J$ large compared to the momentum scale. In the field theory we identify $e^2=J$. 
The renormalization of $e^2$ can be obtained as usual from the parity-even 
contribution to the vacuum polarization, $\Pi(p)$. Thus, the renormalized gauge coupling is given simply by,
\begin{equation}
e_R^2=\frac{e^2}{1+e^2\Pi(0)}. 
\end{equation}
Gauge invariance implies that $\Pi(0)=k/\widehat{m}_R$, where $k$ is some universal constant. Therefore, for $e^2\to\infty$ we obtain 
that $e_R^2\approx\widetilde{m}_R/k$. Equivalently, keeping $e^2$ fixed and approaching the critical point, $\widetilde{m}_R\to 0$ yields 
the same scaling behavior, leading once more to $e_R^2\approx\widetilde{m}_R/k$. Since $e^2$ is identified by the duality as the bare 
phase stiffness, the scaling behavior $e_R^2\approx\widetilde{m}_R/k$ corresponds precisely to the Josepshon scaling 
relation \cite{JOSEPHSON1966608}. At the same time we expect that 
the dimensionless renormalized coupling $\widetilde{\lambda}_R/\widetilde{m}_R$ approaches the (gauged) Wilson-Fisher fixed point as 
$\widetilde{m}_R\to 0$.  

What does the above picture imply for the Dirac fermions? It is usually conjectured that $m_B^2=0$ implies 
$m_0=0$ \cite{Raghu_PhysRevLett.120.016602,SEIBERG2016395,Karch_PhysRevX.6.031043}.    
Inserting this into Eq. (\ref{Eq:mB2-m0}) yields at lowest order based on the mean-field result (\ref{Eq:phi-mean-field}), 
\begin{equation}
\label{Eq:m0-mB=0}
\frac{m_0a}{3}=\ln 2,
\end{equation}
which clearly never vanishes. However, it is important to realize that the actual critical point corresponds to $m_R=0$, which is 
in general 
not attained for $m_B^2=0$.  Furthermore, a more accurate picture should relate $m_0$ to the phase stiffness 
beyond the one-loop result. But then other complications may arise, as the fermions will presumably not be in the free theory regime 
any longer.

\section{Conclusion}

In \cref{sec:pv-duality} we performed in the lattice an exact duality transformation mapping the partition function of bosons in the ACSH model to 
the partition function for an ensemble of closed vortex loop excitations of the same model, which corresponds in field theory language to the 
correspondence shown in the first two lines of the above equation. Furthermore, 
in \cref{sec:comparison} we were able to identify the partition function for an ensemble of vortex loops of the ACSH model 
in the low-energy regime to the partition 
function for an ensemble of closed fermion worldlines. 

More precisely, we have studied in this work the correspondence  between the ACSH model and free massive fermions within the framework of 
a particle-vortex duality. This was achieved 
via an exact duality transformation where closed worldline of bosonic particles arising in the partition function of the ACSH model are brought 
to an equivalent form summing over an ensemble of closed vortex loops of the same model. Thanks to the CS term in the action, this standard 
particle-vortex duality features a phase factor in the partition function where the phase is given by the writhe number of a pair of vortex loops.  
We then showed that the fermionic partition function represented as a sum over an ensemble of closed paths of fermionic particles features 
exactly the same phase factor involving the writhe.  In this case the match between the fermionic and bosonic partition functions is established 
in the low-energy regime where the mass of the fermions is naturally related to the vortex core energy. It turns out that the latter also 
corresponds to the energy density per element of the path of the bosons in the particle representation of the partition function. 

Various aspects of this bosonization duality have been studied intensely in the past, providing conjectures as well as exact results in several limiting cases. Our calculation focus on the self-dual point $\theta=\pi$ of the particle-vortex duality of the ACSH model, and provides exact duality mappings for the lattice versions of  Eqs.~\eqref{Eq:pvduality} and \eqref{eq:fbduality-1}. We find that the bosonization duality holds for $\theta=\pi$ because the 
Gauss linking number contributions are then suppressed from the bosonic partition function, leaving only the writhe number contribution. Concretely, the bosonic dual partition function takes the form of a sum over all possible vortex loops with 
a given writhe yielding the phase factor mentioned above.  

\acknowledgments
This work is supported by the DFG through the W\"urzburg-Dresden Cluster of Excellence on Complexity and Topology in Quantum Matter -- \textit{ct.qmat} (EXC 2147, project-id 39085490) and through SFB 1143 (project-id 247310070). OT and TM furthermore acknowledge financial support by the DFG via the Emmy Noether Programme ME4844/1-1.

\bibliography{arxiv_sub_v1}

\end{document}